\documentclass[aps,superscriptaddress]{revtex4}
\usepackage{epsfig}
\usepackage{graphics,graphicx}
\usepackage{amsmath}
\usepackage{amsbsy}
\usepackage{color}

\begin{document}

\title{Homodyne estimation of quantum states purity by exploiting covariant
uncertainty relation.}
\author{V.I. Man'ko}
\email{manko@na.infn.it}
\affiliation{P.N.Lebedev Physical Institute, Moscow 119991, Russia}
\author{G. Marmo}
\email{marmo@na.infn.it}
\affiliation{Dipartimento di Scienze Fisiche, Universit\`{a} ``Federico II", Via Cintia,
I-80126 Napoli, Italy}
\affiliation{INFN, Sezione di Napoli, Via Cintia, I-80126 Napoli, Italy}
\affiliation{MECENAS, Universit\'{a} ''Federico II'', Napoli, Italy}
\author{A. Porzio}
\email{porzio@na.infn.it}
\affiliation{MECENAS, Universit\'{a} ''Federico II'', Napoli, Italy}
\affiliation{CNR--SPIN, Complesso Universitario di Monte S. Angelo, Via Cintia, I-80126
Napoli, Italy}
\author{S. Solimeno}
\email{solimeno@na.infn.it}
\affiliation{Dipartimento di Scienze Fisiche, Universit\`{a} ``Federico II", Via Cintia,
I-80126 Napoli, Italy}
\affiliation{MECENAS, Universit\'{a} ''Federico II'', Napoli, Italy}
\affiliation{CNISM, Udr Napoli, Via Cintia, I-80126, Napoli, Italy}
\author{F. Ventriglia}
\email{ventriglia@na.infn.it}
\affiliation{Dipartimento di Scienze Fisiche, Universit\`{a} ``Federico II", Via Cintia,
I-80126 Napoli, Italy}
\affiliation{INFN, Sezione di Napoli, Via Cintia, I-80126 Napoli, Italy}
\affiliation{MECENAS, Universit\'{a} ''Federico II'', Napoli, Italy}

\begin{abstract}
We experimentally verify uncertainty relations for mixed states in the
tomographic representation by measuring the radiation field tomograms, \emph{%
i.e.} homodyne distributions. Thermal states of single-mode radiation field
are discussed in details as paradigm of mixed quantum state. By considering
the connection between generalised uncertainty relations and optical
tomograms is seen that the purity of the states can be retrieved by
statistical analysis of the homodyne data. The purity parameter assumes a
relevant role in quantum information where the effective fidelities of
protocols depend critically on the purity of the information carrier states.
In this contest the homodyne detector becomes an easy to handle
purity--meter for the state on--line with a running quantum information
protocol.
\end{abstract}

\maketitle




\baselineskip=19pt%

\section{Introduction}

Quantum pure states can be faithfully described by their wavefunction $\psi
\left( x\right) $. Experimentally producing and measuring pure quantum
states is impossible due to the imperfections both of the generation
procedures and the measurement apparata. In particular quantum states of the
optical fields are always mixed due to impossibility of transmitting and
detecting optical fields with 100\% efficiencies. Density matrix $\hat{\rho}$
formalism encopasses the possibility of describing both pure and mixed
state. In this context an important parameter is represented by the purity $%
\tilde{\pi}$ given by the trace of the squared density matrix operator with $%
\tilde{\pi}=1$ for pure states while $\tilde{\pi}<1$ for mixed ones.
Moreover, $\tilde{\pi}$ assumes a relevant role in quantum information
protocols \cite{vanLock}, where the fidelity, \textit{i.e.} the rate of
success, depends critically on the purity of the states. $\tilde{\pi}$ also
can give a quantitative measure of the decoherence the pure state have
suffered.

Pure states satisfy the Schr\"{o}dinger \cite{Schr30} and Robertson \cite%
{Rob29} basic inequalities that generalize Heisenberg \cite{Heis27}
uncertainty relations including contributions from covariance of conjugate
quantum observables. In Ref. \cite{Dodonov183} (see also \cite{DodPur02}) a
new bound, higher than the Schr\"{o}dinger-Robertson, which accounts for the
contribution of the purity of mixed states was found. In Ref. \cite%
{UncertaintyASL} the Schr\"{o}dinger-Robertson uncertainty relation was
expressed in terms of homodyne tomograms. The generic tomographic approach
to quantum systems was reviewed in \cite{ibort}. Combining the purity
dependent bound and its expression in terms of homodyne tomograms give rise
to a method for a simple determination of the state purity via homodyne
detection developed in \cite%
{Raymer,Mlynek,Lvovski,Solimeno,Bellini,RaymerRMP}.

The problem of measuring the position and momentum was discussed also in
connection with the tomographic approach recently in \cite{Lahti}.

So far, evaluation of purity in continuous variable (CV) systems is obtained
once the full state density matrix have been reconstructed by quantum
tomographic methods. In this paper we prove that the purity of the analyzed
state can be retrieved by a very simple and fast analysis of homodyne data
thus giving the possibility of having an on-line monitor on the state.

The aim of the present paper is twofold. On the one hand we express the
purity $\tilde{\pi}$ of a mixed state in terms of homodyne tomograms, \emph{%
i.e.} quantities amenable to an experimental determination. On the other
hand, relying on the uncertainty relations of Ref. \cite{Dodonov183} we
derive a simple estimator for the purity of a thermal state which is
independent from its tomographic expression.

In Ref. \cite{mancini96} a probability representation of quantum mechanics
was suggested in which states are described by standard probability
distributions, called tomograms. This representation is based on the
representation of the Wigner function $W\left( p,q\right) $ of a quantum
state by means of the Radon integral transform or marginal distributions
(optical tomograms) \cite{Ber-Ber, Vog-Ris}. These data representing the
output of optical homodyne detectors allows reconstructing the Wigner
function using the experimental data. This tomography procedure is nowadays
a routine method for measuring quantum states (see for a review \cite%
{RaymerRMP,ParisSpringer}). In the present paper, using ideas discussed in 
\cite{UncertaintyASL}, we apply the optical tomography to check generalized
Schr\"{o}dinger-Robertson uncertainty relations for conjugate quadratures in
the case of a mixed quantum state, in particular, for measuring the
effective temperature of a state. The idea of our consideration is based on
suggestion \cite{mancini96} (see, also \cite{ibort}) that the homodyne
tomogram is a primary object identified with the quantum state. Due to this,
one can extract all information on the state properties including the purity
from the measured tomogram only, avoiding the reconstruction of the Wigner
function procedure.

The paper is organized as follows.

In section 2, starting from the symplectic forms we give an expression of
the purity parameter and of the mean photon number of any mixed photon state
in terms of measurable optical tomograms. For thermal states this expression
provides the temperature of the field and the mean photon number thus
suggesting to check the accuracy of homodyne detection by comparing photon
statistics obtained in this way, i.e., from optical tomograms, and in
independent photon counting experiment.

In section 3, we review mixed state uncertainty relations to get an
estimation of the purity independent of its tomographic expression, by
saturating the bound set by the uncertainty relation found in \cite%
{Dodonov183}. Hence, measuring optical tomograms of photon states, we can
evaluate the purity of the state, and, in the case of thermal states, study
the dependence of the quantum bound on the field temperature, thus obtaining
a sort of a thermometer for evaluating the temperature. Also, we obtain an
estimation for the mean photon number. Finally, in section 4, an
experimental comparison, for thermal states, is performed between our
estimation and an independent measure of purity of the state, which allows
for evaluating the accuracy of our approximate estimation.

\section{Tomogram as purity and thermal meters}

A quantum state, either mixed or pure, is described by a density operator $%
\hat{\rho}$ with purity parameter $\tilde{\pi}$\ given by:%
\begin{equation*}
\tilde{\pi}=\mathrm{Tr}\left[ \hat{\rho}^{2}\right] \leq 1\ .
\end{equation*}%
According to \cite{mancini96} the state is described by a symplectic
tomogram $\mathcal{W}\left( X,\mu ,\nu \right) ,$ where $X,\mu ,\nu $\ are
real parameters, obtained by the Radon transform of the Wigner function $%
W\left( p,q\right) $ (hereafter\textrm{,}$\mathrm{\ }\hbar =1$):%
\begin{equation}
\mathcal{W}\left( X,\mu ,\nu \right) =\int W\left( p,q\right) \delta \left(
X-\mu q-\nu p\right) \frac{dpdq}{2\pi }=\mathrm{Tr}\left[ \hat{\rho}\ \delta
\left( \hat{X}-\mu \hat{Q}-\nu \hat{P}\right) \right] \text{ ,}\   \label{B}
\end{equation}%
with $\delta \left( \hat{X}-\mu \hat{Q}-\nu \hat{P}\right) $ standing for 
\begin{equation*}
\hat{X}=\mu \hat{Q}+\nu \hat{P}
\end{equation*}%
Accordingly the generalized quadrature operator $\hat{X}\left( \mu ,\nu
\right) $ depends parametrically on $\mu ,\nu $ and its moments are given
by: 
\begin{equation*}
\left\langle \hat{X}^{n}\left( \mu ,\nu \right) \right\rangle =\int X^{n}%
\mathcal{W}\left( X,\mu ,\nu \right) dX\ ,\ n=1,2,3,\ldots
\end{equation*}%
like means and variances, in terms of homodyne quadrature statistics. In
particular, 
\begin{equation}
\hat{X}^{2}\left( \mu ,\nu \right) =\mu ^{2}\hat{Q}^{2}+\nu ^{2}\hat{P}%
^{2}+2\mu \nu \left( \frac{\hat{Q}\hat{P}+\hat{P}\hat{Q}}{2}\right) \text{\ .%
}  \label{E}
\end{equation}%
so that 
\begin{equation}
\left\langle \left( \hat{X}\left( \mu ,\nu \right) -\left\langle \hat{X}%
\left( \mu ,\nu \right) \right\rangle \right) ^{2}\right\rangle =\sigma
_{XX}\left( \mu ,\nu \right) =\mu ^{2}\sigma _{QQ}+\nu ^{2}\sigma _{PP}+2\mu
\nu \sigma _{PQ}  \label{F}
\end{equation}%
Accordingly, the mixed state is characterized by the quadrature dispersions $%
\sigma _{QQ},\sigma _{PP},\sigma _{PQ},$ which in turn can be determined by
measuring $\sigma _{XX}\left( \mu ,\nu \right) $ for particular values of $%
\mu ,\nu .$ In case of optical tomograms, one has $\mu =\cos \theta ,\nu
=\sin \theta $ so that the optical field state is characterized by the field
quadratures relative to $\theta =0,\pi /2$ and $\theta =\pi /4$ : 
\begin{eqnarray*}
\sigma _{QQ} &=&\sigma _{XX}\left( 1,0\right) \ ,\ \sigma _{PP}=\sigma
_{XX}\left( 0,1\right) \ , \\
\sigma _{PQ} &=&\sigma _{XX}\left( \frac{\sqrt{2}}{2},\frac{\sqrt{2}}{2}%
\right) -\frac{1}{2}\left[ \sigma _{XX}\left( 1,0\right) +\sigma _{XX}\left(
0,1\right) \right] \ .
\end{eqnarray*}

Analogously the photon number operator $\hat{n}=\hat{a}^{\dagger }\hat{a}=%
\frac{1}{2}\left( \hat{Q}^{2}+\hat{P}^{2}-1\right) $ is related to the
moments 
\begin{equation}
\left\langle \hat{n}\right\rangle =\frac{1}{2}\left[ \left\langle \hat{X}%
^{2}\left( 1,0\right) \right\rangle +\left\langle \hat{X}^{2}\left(
0,1\right) \right\rangle -1\right] \ .  \label{mean-photon-number}
\end{equation}%
Hence the accuracy of the homodyne detection could be assessed by comparing $%
\left\langle \hat{n}\right\rangle $ obtained via optical tomograms with that
measured by standard photon counting experiments.

The purity $\tilde{\pi}$ of $\hat{\rho}$ is a functional of $\mathcal{W}%
\left( X,\mu ,\nu \right) $ (see, for example, \cite{ibort}) 
\begin{equation*}
\tilde{\pi}=\mathrm{Tr}\left[ \rho ^{2}\right] =\frac{1}{2\pi }\int dXdYd\mu
d\nu \left[ \mathrm{e}^{i\left( X+Y\right) }\mathcal{W}\left( X,\mu ,\nu
\right) \mathcal{W}\left( Y,-\mu ,-\nu \right) \right]
\end{equation*}%
or equivalently 
\begin{equation}
\tilde{\pi}=\frac{1}{2\pi }\iint dXdY\int\nolimits_{0}^{2\pi }d\theta
\int\nolimits_{0}^{\infty }dk\left[ k\mathrm{e}^{ik\left( X+Y\right) }%
\mathcal{W}_{0}\left( X,\theta \right) \mathcal{W}_{0}\left( Y,\theta +\pi
\right) \right] \text{\ ,}  \label{OptPur}
\end{equation}%
having replaced $\mathcal{W}\left( X,\mu ,\nu \right) $ with the homodyne
marginal distribution 
\begin{equation*}
\mathcal{W}_{0}\left( X,\theta \right) =k\mathcal{W}\left( kX,k\cos \theta
,k\sin \theta \right)
\end{equation*}%
which is accessed in homodyne measurements.

For Gaussian photon states $\mathcal{W}\left( X,\mu ,\nu \right) $ reduces
to: 
\begin{equation}
\mathcal{W}(X,\mu ,\nu )=\frac{1}{\sqrt{2\pi \sigma _{XX}\left( \mu ,\nu
\right) }}\exp \left[ -\frac{\left( X\left( \mu ,\nu \right) -\left\langle
X\left( \mu ,\nu \right) \right\rangle \right) ^{2}}{2\sigma _{XX}\left( \mu
,\nu \right) }\right]   \label{GauTom}
\end{equation}%
which inserted (\ref{GauTom}) into (\ref{OptPur}) yields the well-known
expression: 
\begin{equation}
\tilde{\pi}=\frac{1}{2\sqrt{\sigma _{QQ}\sigma _{PP}-\sigma _{QP}^{2}}}\ .
\label{purth}
\end{equation}%
For a thermal state $\left( \sigma _{QQ}=\sigma _{PP}=\frac{1}{2}\coth
\left( \frac{1}{2T}\right) \ ;\ \sigma _{QP}=0\right) $ $\tilde{\pi}$
reduces to 
\begin{equation}
\tilde{\pi}=\tanh \left( \frac{1}{2T}\right)   \label{TemPur}
\end{equation}%
with $T$ measured in $Kelvin\times K_{B}/\hbar \omega $.

\section{Mixed state uncertainty relation}

\begin{figure}[h]
\centering
\includegraphics[scale=0.5]{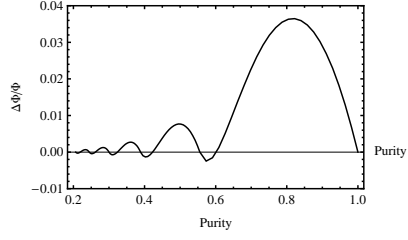}
\caption{Relative difference between the bound $\Phi ^{2}$ and its
approximation $\widetilde{\Phi }^{2}$ given in Eq. (\protect\ref{phi_approx}%
). $\widetilde{\Phi }^{2}$ approximates $\Phi ^{2}$ within a few percents.}
\label{plotdeltaFI}
\end{figure}

The general uncertainty relation for a mixed state reads \cite{Dodonov183} 
\begin{equation}
\sigma _{QQ}\sigma _{PP}-\sigma _{QP}^{2}\geq \frac{1}{4}\Phi ^{2}(\tilde{\pi%
}).  \label{pos-mom covar}
\end{equation}%
with $\tilde{\pi}$ the purity of the state $\hat{\rho}$. The real,
continuous and differentiable function $\Phi (\tilde{\pi}),$ such that $\Phi
(\tilde{\pi})\geq 1$ in the interval $0<\tilde{\pi}\leq 1$, has the
following piecewise analytic expression (extrema of intervals are given by $%
2(2k+1)/3k(k+1)$, $k=1,2,\ldots $): 
\begin{eqnarray}
\Phi (\tilde{\pi}) &=&2-\sqrt{2\tilde{\pi}-1}\quad \quad \frac{5}{9}\leq 
\tilde{\pi}\leq 1  \notag \\
\Phi (\tilde{\pi}) &=&3-\sqrt{8\left( \tilde{\pi}-\frac{1}{3}\right) }\quad
\quad \frac{7}{18}\leq \tilde{\pi}\leq \frac{5}{9}  \notag \\
\Phi (\tilde{\pi}) &=&4-\sqrt{20\left( \tilde{\pi}-\frac{1}{4}\right) }\quad
\quad \frac{3}{10}\leq \tilde{\pi}\leq \frac{7}{18} \\
&\cdots \cdots \cdots &  \notag
\end{eqnarray}%
Besides, the function $\Phi (\tilde{\pi})$ can be approximated in the whole
interval $(0,1)$ within $1\%$ by the interpolating function \cite%
{Dodonov183, DodPur02}: 
\begin{equation}
\widetilde{\Phi }(\tilde{\pi})=\frac{4+\sqrt{16+9\tilde{\pi}^{2}}}{9\tilde{%
\pi}}.  \label{phi_approx}
\end{equation}%
In Fig. (\ref{plotdeltaFI}) we plot the relative difference between $\Phi
^{2}(\tilde{\pi})$, the square of the bound function, and $\widetilde{\Phi }%
^{2}(\tilde{\pi})$ in order to visualize how good is the approximation. 

We generalize the inequality (\ref{pos-mom covar}) to arbitrary local
oscillator phases $\theta $, by using the tomographic uncertainty function $%
F(\theta )$ introduced in \cite{UncertaintyASL}: 
\begin{equation}
F(\theta )\geq \frac{1}{4}\left[ \Phi ^{2}(\tilde{\pi})-1\right] .
\label{ManDodGen}
\end{equation}

Then, by using $\widetilde{\Phi }(\tilde{\pi})$ instead of $\Phi (\tilde{\pi}%
)$, the uncertainty relation reads: 
\begin{equation}
F\left( \theta \right) -\frac{1}{4}\left[ \widetilde{\Phi }^{2}(\tilde{\pi}%
)-1\right] =F\left( \theta \right) -\left[ \frac{8+2\sqrt{16+9\tilde{\pi}^{2}%
}-18\tilde{\pi}^{2}}{81\tilde{\pi}^{2}}\right] \geq 0\ ,  \label{ManDod}
\end{equation}
in this way, a relation between the tomographic uncertainty function $%
F\left( \theta \right) $ and the state purity is set.

We recall that $F\left( \theta \right) $ is defined, by means of the
variance of the homodyne quadrature $\hat{X}$ 
\begin{equation}
\sigma _{XX}\left( \theta \right) =\int X^{2}\mathcal{W}_{0}(X,\theta )dX-%
\left[ \int X\mathcal{W}_{0}(X,\theta )dX\right] ^{2}\ ,
\end{equation}%
as: 
\begin{eqnarray}
F(\theta ):= &&\sigma _{XX}\left( \theta \right) \sigma _{XX}\left( \theta +%
\frac{\pi }{2}\right) +  \notag \\
&&-\left[ \sigma _{XX}\left( \theta +\frac{\pi }{4}\right) -\frac{1}{2}%
\left( \sigma _{XX}\left( \theta \right) +\sigma _{XX}\left( \theta +\frac{%
\pi }{2}\right) \right) \right] ^{2}-\frac{1}{4}.  \label{F(Theta)}
\end{eqnarray}

We note that for $\theta =0$ one has 
\begin{equation}
\left. F\left( \theta \right) \right\vert _{\theta =0}=\sigma _{QQ}\sigma
_{PP}-\sigma _{QP}^{2}-\frac{1}{4}\ .  \label{S-R}
\end{equation}%
So that, $\left. F\left( \theta \right) \right\vert _{\theta =0}\geq 0$ is
exactly the Schr\"{o}dinger-Robertson inequality.\ Moreover, comparing the
last expression with the thermal state purity given in Eq. (\ref{purth}) it
is easy to see that:%
\begin{equation*}
\tilde{\pi}=\frac{1}{2\sqrt{\left. F\left( \theta \right) \right\vert
_{\theta =0}+\frac{1}{4}}}~.
\end{equation*}
The above expression of the uncertainty function $F\left( \theta \right) $
in terms of tomograms was given in \cite{UncertaintyASL}.

The physical meaning of the function $F\left( \theta \right) $ is the
following. For a local oscillator phase $\theta =0$, it is the determinant
of the quadrature dispersion matrix, shifted by $-1/4$ as shown in Eq. (\ref%
{S-R}). For a nonzero local oscillator phase $\theta $, the function $%
F\left( \theta \right) $ corresponds to the determinant of the dispersion
matrix of the quadratures, which are measured in a rotated reference frame
in the quadrature phase space. The non-negativity of the function $F\left(
\theta \right) $ implies the fulfilling of the Schr\"{o}dinger-Robertson
uncertainty relation for all the unitarily equivalent position and momentum
operators, since the unitary transformations do not change the canonical
commutation relations. The formula (\ref{F(Theta)}) simply expresses the
determinant of the dispersion matrix for unitarily rotated position and
momentum, in terms of tomographic probability distribution $\mathcal{W}%
_{0}(X,\theta )$.

\begin{figure}[ht]
\centering
\includegraphics[scale=0.7]{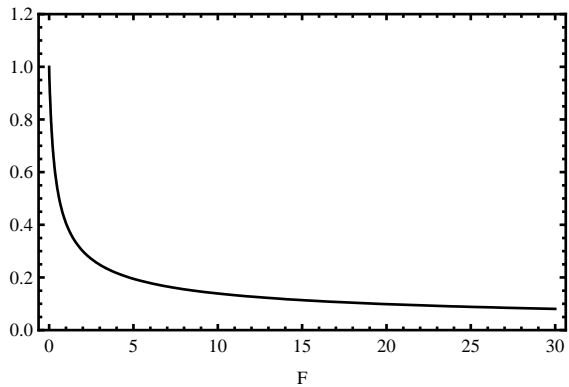}
\caption{Purity vs. $F$ the minimum value of $F\left( \protect\theta \right) 
$. $F$ saturates the inequality (\protect\ref{ManDod}).}
\label{plotpur}
\end{figure}

Thus, the tomogram must satisfy the inequality (\ref{ManDodGen}), or (\ref%
{ManDod}), where the parameter $\tilde{\pi}$ is expressed in tomographic
terms by Eq. (\ref{OptPur}). However, we can get an estimation of the purity
in terms of the tomographic uncertainty function $F\left( \theta \right) $
by saturating the inequality (\ref{ManDod}). In other words, we consider the
minimum value $F$ of the uncertainty function $F\left( \theta \right) $ and
estimate that for such a value the inequality is pretty near saturated.
Then, by solving with respect to $\tilde{\pi}$, we are able to express the
purity as a function of $F$. This is a simple expression: 
\begin{equation}
\tilde{\pi}\left( F\right) \approx \frac{2\sqrt{1+4F}}{2+9F}\text{\ ,}
\label{PurApprox}
\end{equation}%
whose plot is shown in Fig. \ref{plotpur}. 

As one can see, $F=0$ corresponds to a pure state, $\tilde{\pi}=1$, and the
purity is a smooth decreasing function, going to zero for $F$ at infinity.

On the other hand, for thermal states the purity parameter is directly
related to the temperature, i.e. $\tilde{\pi}=\tanh (1/2T)$, so that the
previous inequalities and estimations can be translated in terms of
temperature. So, in particular, for a thermal state the bound $\tilde{\Phi}$
in inequality (\ref{ManDod}) can be written in terms of the temperature $T$.

Moreover, $F\left( \theta \right) $ provides the temperature $T\left(
F\right) $ of a thermal state with the corresponding purity $\tilde{\pi}%
\left( F\right) $. We get%
\begin{equation}
T=\left[ 2\tanh ^{-1}\tilde{\pi}\right] ^{-1}\approx \left[ 2\tanh
^{-1}\left( \frac{2\sqrt{1+4F}}{2+9F}\right) \right] ^{-1}\ .  \label{TofF}
\end{equation}

We recall that for a thermal state the tomographic uncertainty function $%
F(\theta )$ does not depend on the local oscillator phase $\theta $, as
shown in \cite{UncertaintyASL}, so that $F(\theta )=F,\ \forall \theta $.

Finally, we remark that for a thermal state the mean value of the photon
number can be expressed as a function of the temperature as 
\begin{equation}
\left\langle \hat{n}\right\rangle =\frac{1}{2}\coth \left( \frac{1}{2T}%
\right) -\frac{1}{2}\ ,
\end{equation}%
so we have a relation between our estimation of the temperature $T\left(
F\right) $ and the photon statistics.

In other words, we may check the accuracy of our estimation of the purity
and the temperature, resulting from inequality (\ref{ManDod}), by comparing
our estimation of $\left\langle \hat{n}\right\rangle $, i.e. 
\begin{equation}
\left\langle \hat{n}\right\rangle \left( F\right) =\frac{2+9F}{4\sqrt{1+4F}}-%
\frac{1}{2}\text{\ ,}  \label{nofF}
\end{equation}%
with an independent measure of the mean number of photons. This comparison
is discussed in the next section.

We conclude this section by observing that if the state is not thermal, but
we succeed in finding the value $F$ saturating the uncertainty inequality to
introduce an effective temperature $T_{eff}.$ This effective temperature is
again given by the right hand side of Eq. (\ref{TofF}) and corresponds to a
purity parameter $\tilde{\pi}\left( F\right) $.

\section{Experiment}

Being the uncertainty relations related to optical tomograms through the
tomographic function $F\left( \theta \right) $ (see Eq. (\ref{ManDodGen}))
experimental data obtained by optical homodyne detector, suitable for
retrieving $F\left( \theta \right) $, allow checking the uncertainty
relation (\ref{ManDod}). Moreover, $F\left( \theta \right) $ allows to
evaluate $\tilde{\pi}$, the state purity (Eq. (\ref{PurApprox})), and in the
case of thermal state, $T$, the field temperature (Eq. (\ref{TofF})), and $%
\left\langle \hat{n}\right\rangle $, the mean photon number (Eq. (\ref{nofF}%
)).

Thermal states are Gaussian so for these states also Eq. (\ref{purth}) is
valid. Thus, in order to asses the reliability of the proposed method, it is
possible to compare the estimations of $\tilde{\pi}$ via $F\left( \theta
\right) $ with the same quantity obtained by using Eq. (\ref{purth}).
Moreover, a full reconstruction of the state via quantum tomography provides
a further estimation of $\tilde{\pi}$.

To this end, $F\left( \theta \right) $ has been retrieved for thermal
continuous wave (CW) states, outing a sub-threshold non-degenerate optical
parametric oscillator (OPO) \cite{Dauria08}. In such a device non--linear
fluorescence gives rise to a pair of down-converted entangled modes each in
a thermal state \cite{JOSAB2010}. The experimental setup, illustrated in
greater details elsewhere \cite{Dauria08,JOSAB2010}, can be sketched into
three distinct blocks: the state source, the below threshold OPO, the
detector, a quantum homodyne, and the acquisition board.

The quantum homodyne detector, shows an overall quantum efficiency $\eta
=0.88\pm 0.02$ (see Refs. \cite{Dauria05,Dauria06,Solimeno} for details).
The system is set to obtain a $2\pi $--wide linear scanning of the LO phase
in an acquisition window. Since $F\left( \theta \right) $ is retrieved by
combining variances of data distributions calculated at different $\theta $,
we have decided to retrieve $F\left( \theta \right) $ in $[0,\pi ]$.

In order to use our homodyne data to calculate $F\left( \theta \right) $, we
must be sure that the state is effectively a thermal one. First we prove
that the state is Gaussian by performing some tests to assess the
Gaussianity of data distribution \cite{cechi09}. In particular, we have used
the kurtosis excess (or Fisher's index) and the Shapiro-Wilk indicator. The
kurtosis is a measure of the "peakedness" of the probability distribution of
a real-valued random variable while the Shapiro-Wilk one tests the null
hypothesis that a distribution $x_{1},...x_{n}$ came from a normally
distributed population. Then, a pattern function tomographic analysis is
used for assuring the thermal character of the state \cite{Pattern,Solimeno}
and for reconstructing the state Wigner function.

$F\left( \theta \right) $ is calculated by analyzing the data distributions
(each distribution containing 2100 data) at $47$ different values of $\theta 
$ (each $\theta $ value then corresponds to a phase interval interval of $%
0.067$ \mbox{rad}) and making use of Eq. (\ref{F(Theta)}).

\begin{figure}[tb]
\begin{center}
\centering\includegraphics[scale=0.3]{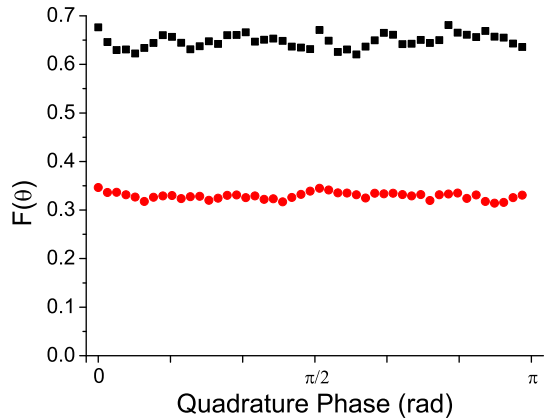}
\end{center}
\caption{$F(\protect\theta )$ vs LO phase $\protect\theta \in \lbrack 0,%
\protect\pi \lbrack $. The mean values are: $0.65\pm 0.01$ (squares, $\tilde{%
\protect\pi}=0.48$), and $0.329\pm 0.007$ (circles, $\tilde{\protect\pi}=0.61
$) where the errors are the standard deviations of the $F\left( \protect%
\theta \right) $ distributions in $\protect\theta .$ The plotted range
corresponds to 47 phase bins (see text for details).}
\label{explot}
\end{figure}

A typical output is reported in Fig. \ref{explot}. The traces are obtained
by subtracting to each experimental value of $F\left( \theta \right) $, $%
\bar{F}_{shot}$ corresponding to the average of $F\left( \theta \right) $
over $\theta $ for a vacuum state whose data are collected by obscuring the
homodyne input. $F\left( \theta \right) $ for a shot noise trace returns the 
$0$ of the instrument. Furthermore, in order to avoid any influence on the
statistics of the data, the electronic noise is kept $\approx 15$ dB below
the shot noise. The values of $F\left( \theta \right) $, always positive as
predicted by the uncertainty relation (\ref{ManDod}), are with a good
approximation independent of $\theta $ as expected for a thermal state.

We have analyzed 218 homodyne acquisitions of thermal states. This large
number allows a statistical approach for analyzing the reliability of the
proposed method. In particular the purity $\tilde{\pi}$ of the state has
been evaluated: a) by reconstructing the state via homodyne tomography ($%
\tilde{\pi}_{tom}$); b) by using Eq. (\ref{PurApprox}) ($\tilde{\pi}_{F}$);
c) by using the exact expression for a thermal state (see Eq. (\ref{purth}))
($\tilde{\pi}_{th}$). 
\begin{figure}[tb]
\begin{center}
\centering\includegraphics[scale=0.4]{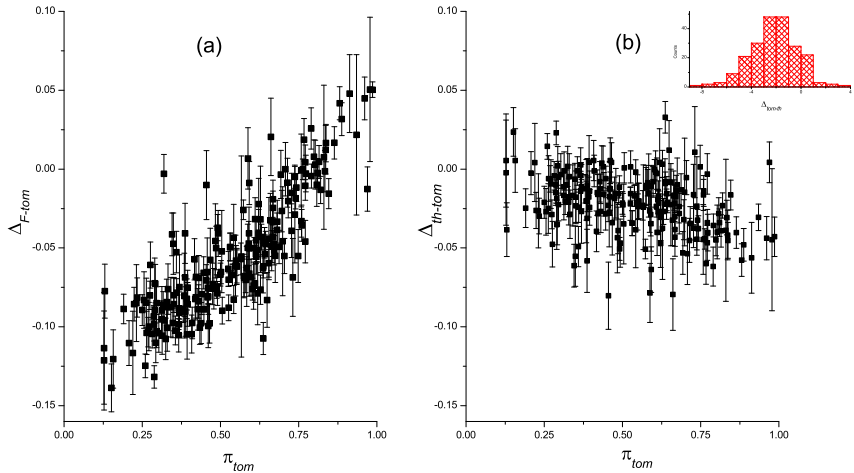}
\end{center}
\caption{$\Delta _{F-tom}$ (a) and $\Delta _{th-tom}$ (b) vs. $\tilde{%
\protect\pi}_{tom}$. The histogram $\Delta _{th-tom}$, given in the inset of
(b), proves that $\Delta _{th-tom}$ is normally distributed. $\Delta _{F-tom}
$ shows a well defined behaviour in $\tilde{\protect\pi}_{tom}$ thus
signalling the insorgence of systemathic error due to the used
approximation.\ The estimation become more precise as the state approaches a
pure one. The maximum error is arround 15\%.}
\label{stat}
\end{figure}
In Fig. \ref{stat} we report the values of the differences $\Delta _{F-tom}$
(a) and $\Delta _{th-tom}$ (b) between $\tilde{\pi}_{F}$ and $\tilde{\pi}%
_{tom}$ and $\tilde{\pi}_{th}$ and $\tilde{\pi}_{tom}$ normalized to their
average. $\tilde{\pi}_{tom}$ has been taken as a reference \footnote{%
The tomographic value is the one giving the lower error due the very high
statistical reliability of such quantum state reconstruction method.}. By
looking at these distributions it is seen that while $\Delta _{th-tom}$ is
normally distributed (see inset in Fig. \ref{stat} (b)) with an average of
0.022 (standard deviation 0.018) $\Delta _{F-tom}$ presents a sistemathic
behavior in $\tilde{\pi}$ thus signalling the onset of systemathic error in
the obtained determination. While $\tilde{\pi}_{th}$ comes from the exact
expression of Eq. (\ref{purth}), $\tilde{\pi}_{F}$ is obtained by
approximating $\phi \left( \tilde{\pi}\right) $ (see Eq. (\ref{phi_approx}))
so that the relative precision depends on the purity itself. It has to be
noted that in any case the maximum relative error remains below $15\%$.

\begin{figure}[tb]
\begin{center}
\centering\includegraphics[scale=0.3]{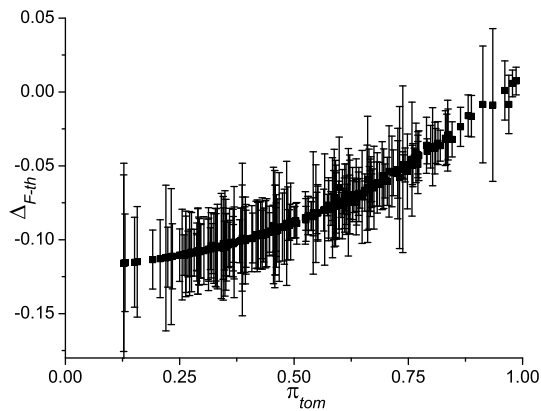}
\end{center}
\caption{$\Delta _{F-th}$ vs. $\tilde{\protect\pi}_{tom}$. presents a
systemathic behaviour in $\tilde{\protect\pi}_{tom}$ due to the
approximation of Eq. (\protect\ref{phi_approx}).}
\label{stat1}
\end{figure}

Comparing $\tilde{\pi}_{F}$ and $\tilde{\pi}_{th}$ evidenziate this effect
even better (see Fig. \ref{stat1}).

The above discussed discrepancies, even not so small, confirms that the
estimation of $F\left( \theta \right) $, much more simpler than any full
state reconstruction methods \cite{RaymerRMP} gives many reliable
information on the homodyned state. Moreover, for a thermal state it is
possible to obtain more precise information by exploiting Eq. (\ref{purth}).

By exploiting the thermal character of the states herein analysed it is
possible to get information on $T$ (Eq. (\ref{TofF})) and $\left\langle \hat{%
n}\right\rangle $ averaging the experimental trace for $F\left( \theta
\right) $ over $\theta $. The values so obtained are given in the following
table (together with their confidence interval): 
\begin{equation}
\begin{tabular}{|c|c|c|c|}
\hline
$\text{mode}$ & $F$ & $T\frac{k_{B}}{\hbar \omega }$ & $\left\langle \hat{n}%
\right\rangle $ \\ \hline
1 & $0.65\pm 0.01$ & $3.8\pm 0.1$ & $0.53\pm 0.09$ \\ \hline
2 & $0.329\pm 0.007$ & $2.8\pm 0.1$ & $0.315\pm 0.006$ \\ \hline
\end{tabular}
\label{results}
\end{equation}

The analyzed mode has $\nu \approx 3\times 10^{14}$ $Hz$ so that the field
temperature is of the order $\approx 10^{4}$ $\mathrm{K}$.

\section{Conclusions}

Uncertainty relations for mixed quantum states \cite{Dodonov183,DodPur02}
contain information that allows to easily estimate the purity once a
connection between optical tomograms and these relations is set. In this
paper it is shown that quadrature statistics of the field contains complete
information on bound in the generalized uncertainty relations of mixed state
and that this bound is intimately related to the state purity. This approach
gives the possibility of a fast and reliable estimation of the purity
parameter for quantum states that does not require sophysticate data
analysis. As far as we know, all the methods for purity estimation rely on
the reconstruction of the whole state density matrix while the approach
presented in this paper allows to recover the purity simply calculating the
tomographic function $F\left( \theta \right) $. The experimental results
herein presented show that the homodyne detector can be used as the purity
meter of the electromagnetic radiation in the quantum domain. The
experimental estimations obtained via $F\left( \theta \right) $ are
compatible with more sophysticate estimation obtained via pattern function
tomography so proving that our method can be used for real--time evaluations
of some basic properties of the homodyne optical states.

Further study of other photon states like squeezed states and multi-mode
states will give the possibility to check other quantum phenomena as higher
momenta uncertainty relations.

\textbf{Acknowledgements} V.I. Man'ko thanks the University `Federico II'
and the Sezione INFN di Napoli for kind hospitality and the Russian
Foundation for Basic Research for partial support under Projects No.
09-02-00142.

\end{document}